\documentclass{iau}
\usepackage{graphicx}

\title[Galactic spins in the cosmic web] 
{How do galaxies build up their spin in the cosmic web?}

\author[Charlotte Welker \& Yohan Dubois]   
{Charlotte Welker$^1,^2$
 Yohan Dubois$^1,^2$  Christophe Pichon$^1$  Julien Devriendt$^2$   \and  Sebastien Peirani$^1$}

\affiliation{$^1$Institut d'Astrophysique de Paris, \\ 98bis boulevard Arago, 75014 Paris, France\\ email: {\tt welker@iap.fr} \\[\affilskip]
$^2$Sub-department of Astrophysics, University of Oxford, \\ Keble Road, Oxford OX1 3RH, United Kingdom.}

\pubyear{2014}
\volume{308}  
\pagerange{119--126}
\jname{The Z'eldovich Universe}
\editors{A.C. Editor, B.D. Editor \& C.E. Editor, eds.}
\begin{document}

\maketitle

\begin{abstract}
Using the Horizon-AGN simulation we find a mass dependent spin orientation trend for galaxies: the spin of low-mass, rotation-dominated, blue, star-forming galaxies are preferentially aligned with their closest filament, whereas high-mass, velocity dispersion- supported, red quiescent galaxies tend to possess a spin perpendicular to these filaments. We explore the physical mechanisms driving galactic spin swings and quantify how much mergers and smooth accretion re-orient them  relative to  their host  filaments. 
\end{abstract}

\firstsection 
\section{Introduction: from haloes to galaxies}

	Over the past ten years, numerous  simulations have reported the influence of the cosmic web on the direction of the angular momentum (AM) of halos. Several numerical investigations (see  \cite[Dubois et al. (2014) for references] )) argued that the spin of high-mass halos tends to lie perpendicular to their host filament, whereas low-mass halos have a spin preferentially aligned with it. This was confirmed with high degree of accuracy by \cite[Codis et al. (2012)] , which quantified a redshift-dependent mass transition $M_{tr,h}$ separating aligned and perpendicular halos.
	They interpreted the origin of the transition in terms of large-scale cosmic flows: high-mass haloes would have spins perpendicular to the filament because they are the results of major mergers while low-mass haloes  acquire their mass by smooth accretion from the vorticity quadrant they are embedded in, resulting in spins parallel to the filament (see also \cite[Laigle et al. 2014])).

 	In order to make predictions for  observational estimators of such correlations, it is important to identify the expected corresponding correlations between {\sl galaxies} and the cosmic web. In this work, we use the {\tt Horizon-AGN} simulation to focus on the influence of the cosmic web as an anisotropic vector of the gas mass and angular momentum which ultimately shapes galaxies. Our  purpose is to determine if the mass-dependent halo spin-filament correlations of \cite[Codis et al. (2012)]{}  can be traced through the morphology and physical properties of simulated galaxies. 

We  first present the method used to identify the effect of the environmentally-driven spin acquisition on morphology, and probe the tendency for galaxies to align or misalign with the cosmic filaments as a function of galactic properties. 
We then focus on explaining {\sl why} galactic spins swing and investigate the competitive effects of mergers and smooth accretion on alignment.

\section{Tracing galactic alignments}

{\underline{\it Horizon-AGN}}

Horizon-AGN is a state-of-the-art hydrodynamical cosmological simulation run on a $L_{box}=100 Mpc.h-1$ box down to z=1.2 with the AMR code RAMSES  ( \cite[Teyssier (2002)]), an initial resolution of $1024^{3}$ dark matter particles and 7 levels of refinement. It assumes a $ \Lambda$CDM cosmology and includes various physical processes like feedback from supernovae and AGN, star formation and cooling from H and He with a contribution from metals. (see details in \cite[Dubois et al (2014)].) With more than 150 000 galaxies and 300 000 haloes per snapshot, it  displays a vast diversity of galaxies, and spans a wide range of galactic morphologies.

{\underline{\it Spin orientation distribution for galaxies}} 

\begin{figure}[b]
\begin{center}
   \includegraphics[width=3.3cm]{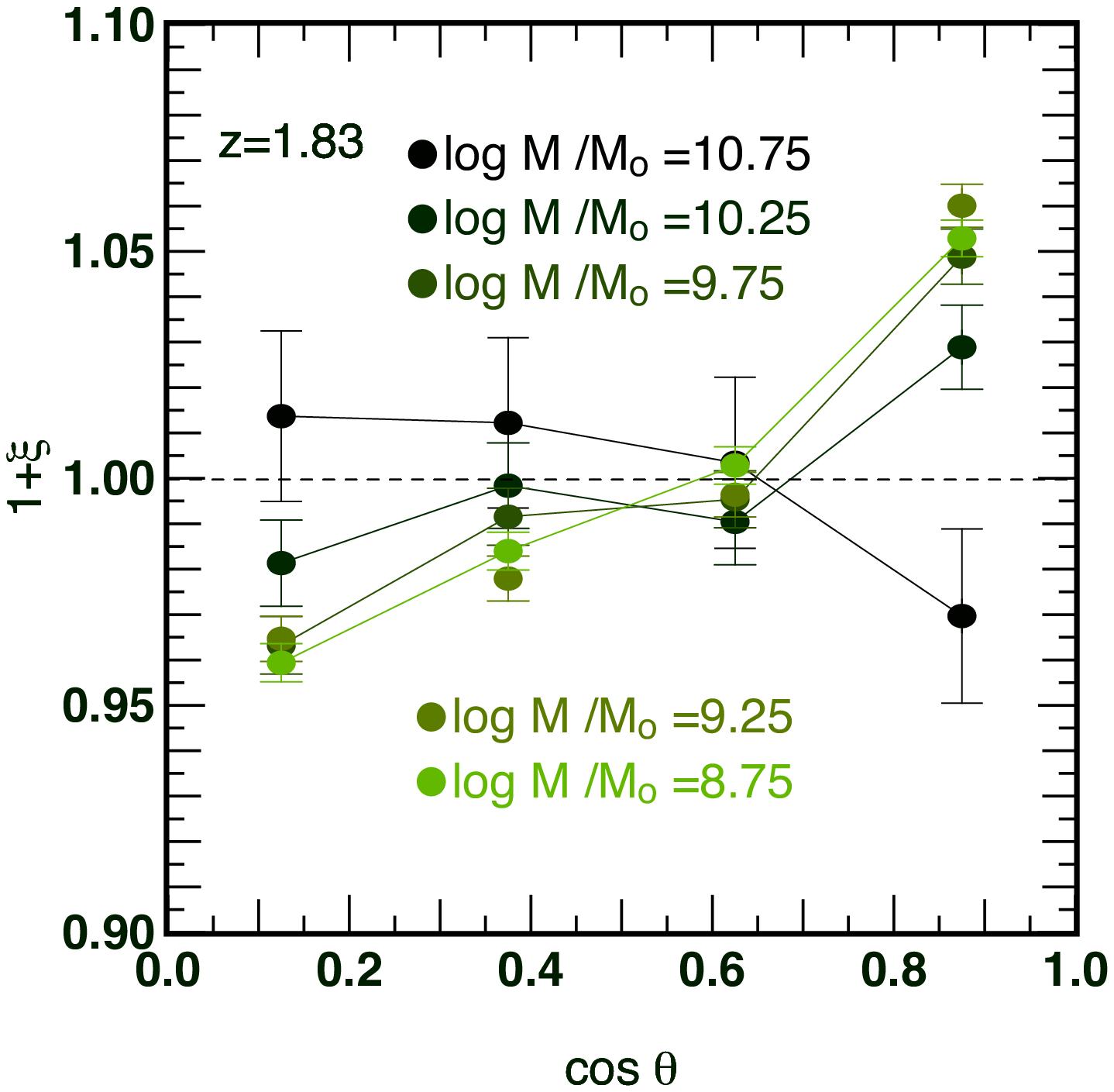}   
      \includegraphics[width=3.3cm]{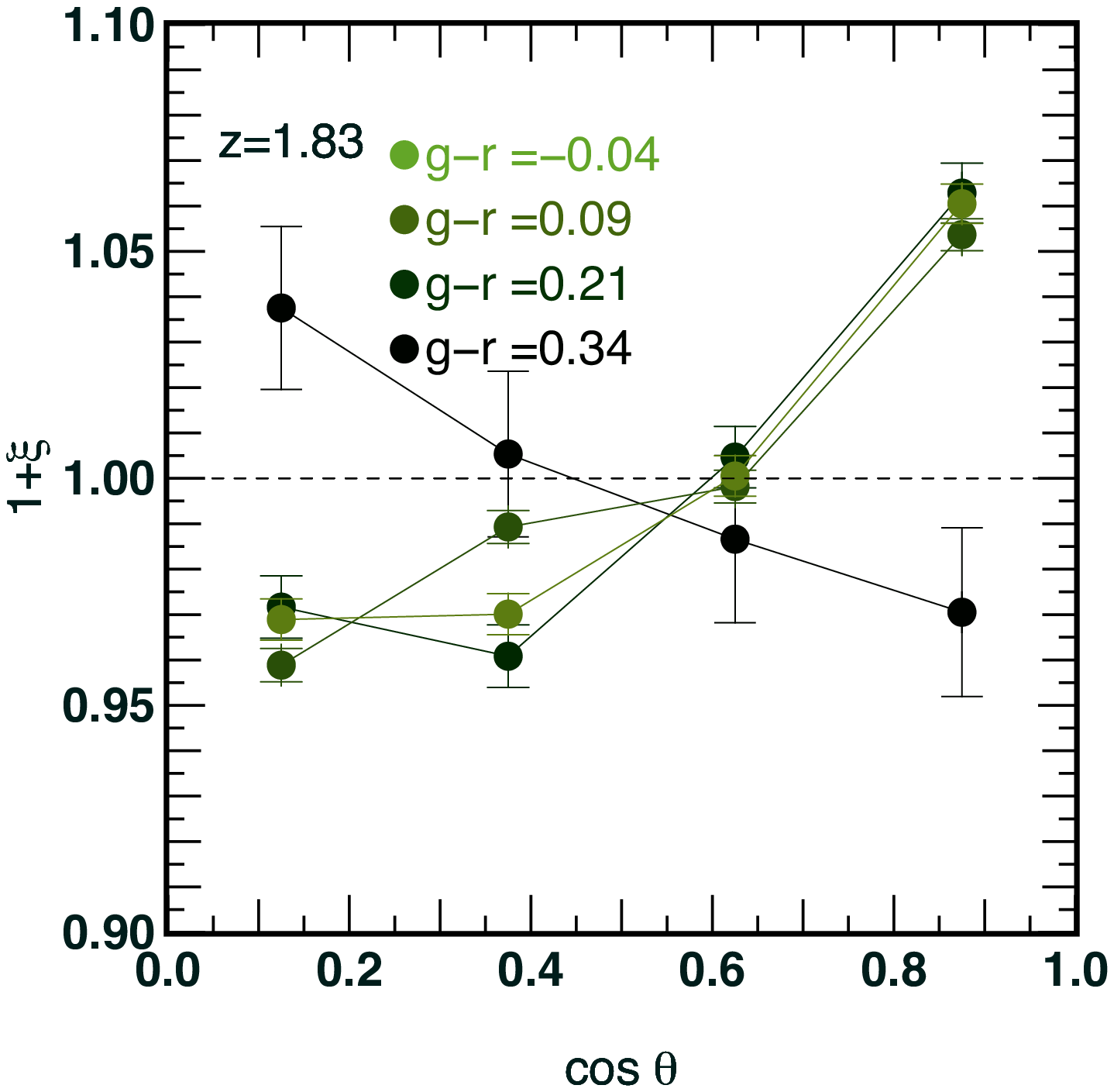}
       \includegraphics[width=3.3cm]{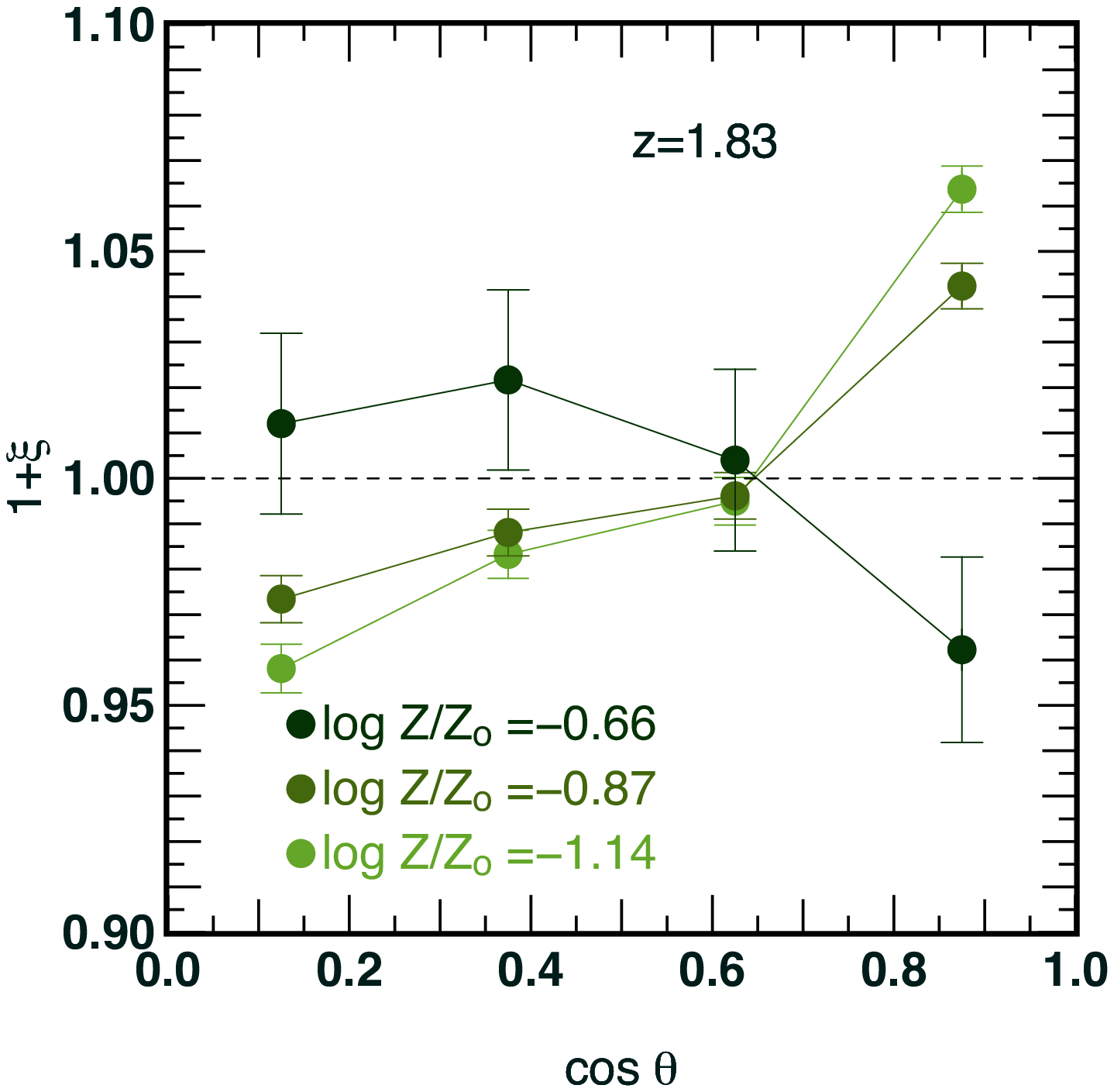}
          \includegraphics[width=3.3cm]{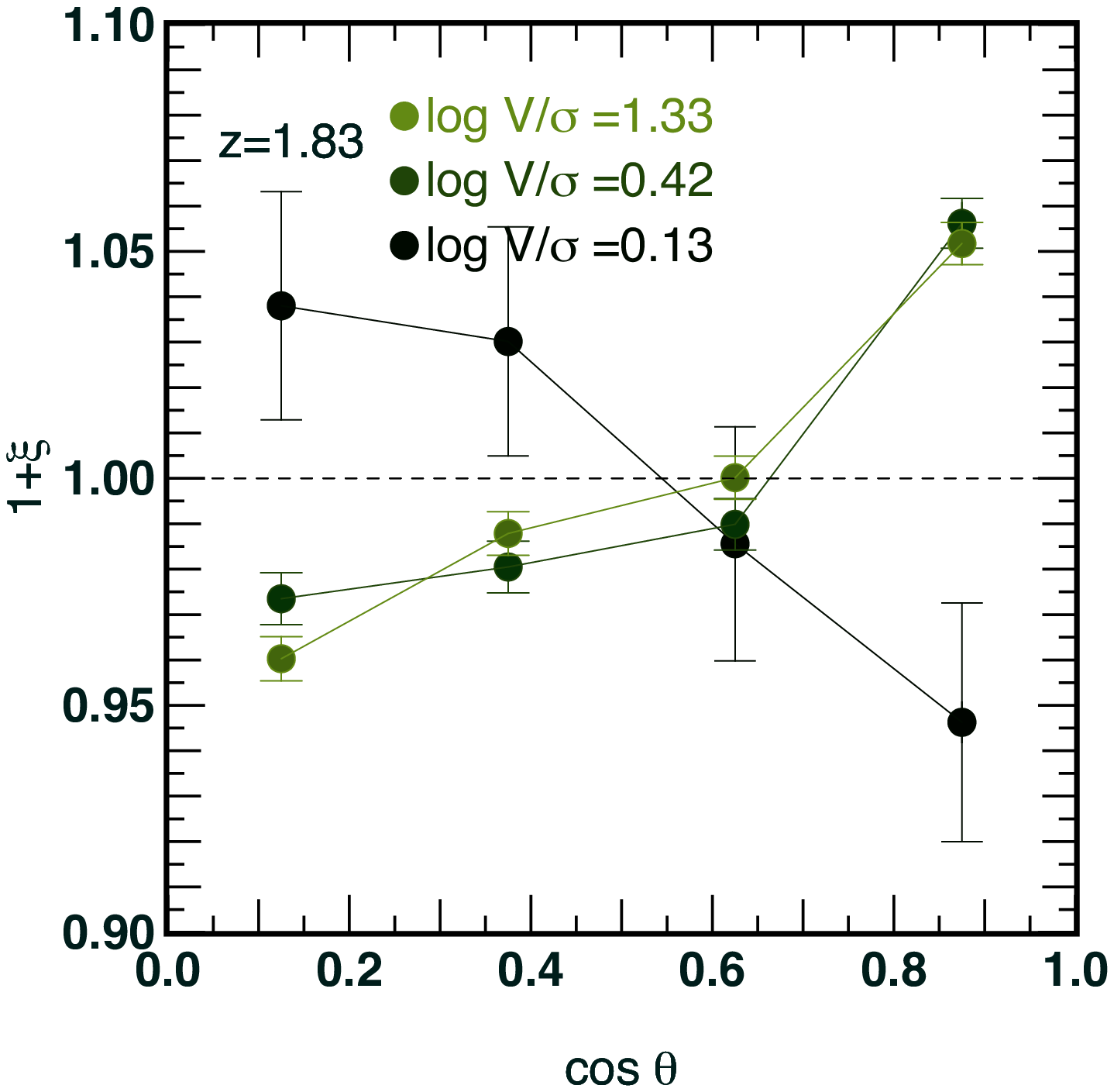}  
 \caption{Excess probability, $\xi$, of the alignment between the spin of galaxies and their closest filament  as a function of galactic properties at $z=1.83$: (from left to right) $M_{\rm s}$, $g-r$, metallicity $Z$ and  $V/\sigma$.
Half sigma error bars are shown for readability. Dashed line corresponds to a uniform PDF (excess probability $\xi=0$).}
   \label{fig1}
\end{center}
\end{figure}

As can be seen in Fig.~\ref{fig1}, we  recover the spin alignment  for small galaxies (and perpendicular spin orientation for massive galaxies) consistent with what is already known for dark haloes. Moreover, we find that various galactic properties traced this transition such as colour, age, specific stellar formation rate or metallicity. In a nutshell, low-mass, young, centrifugally supported, metal-poor, bluer galaxies tend to have their spin aligned to the closest filament, while massive, high velocity dispersion, red, metal-rich, old galaxies are more likely to have a spin perpendicular to it (see \cite[Dubois et al 2014] )). The mass-transition, confidently bracketed between $ log(M_{\rm s}/M_\odot)=10.25$ and $ log(M_{\rm s}/M_\odot)=10.75$, was also found to be consistent with the transition mass for dark haloes estimated in \cite[Codis et al (2012)] .

{\underline{\it Morphology tracers and observations}}

Let us consider in particular the $V/\sigma$ kinematic parameter, with $V$ the luminosity weighted rotation component of the stellar speeds and $\sigma$ the dispersion speed of the stars (see Fig.~\ref{fig1} ). Using the corresponding projected quantities, this pseudo-observable can be used as a morphology tracer. Indeed, $V/\sigma<1$ corresponds to dispersion supported ellipticals while $V/\sigma>1$ traces rotation supported spirals. As expected, we find that spirals tend to have spin parallel to the closest filament while the spin of ellipticals tend to be perpendicular.
This  result allows for direct comparison with recent observations in the SDSS. Indeed \cite[Tempel and Libeskind (2013)]{}  found a similar trend, giving this numerical distribution some observational support. 


The underlying explanation  for  these  trends in various galactic features lies in the fact that they are highly correlated to the corresponding stellar mass
and with the AM distribution within the galaxy. This calls for a dynamical scenario, which we develop now.

\section{Spin swing dynamics}

{\underline{\it Merger trees}}

To understand this morphological segregation, we perform a statistical analysis on 22 outputs equally spaced in redshift between z=5.2 and z=1.2, which loosely corresponds to an average time step of 250 Myr. A candidate for merger being defined as any  structure with $M_{s}>10^8  M_{\odot}$, we then build merger trees to track back in redshift the most massive progenitor of each galaxy at any time step. We define the merger mass ratio  $\delta m=\Delta m_{\rm mer}(z_n)/M_{s}(z_n)$ with $M_{s}(z_n)$ the total stellar mass of a galaxy at redshift $z_n$ and  $\Delta m_{\rm mer}(z_n)$ the stellar mass accreted through mergers between redshifts $z_{n-1}$ and $ z_n$. In a similar spirit, we also quantify the evolution of the specific angular momentum evolution.


{\underline{\it Spin alignments and smooth accretion}}
\begin{figure}[b]
\begin{center}
   \includegraphics[width=3.7cm]{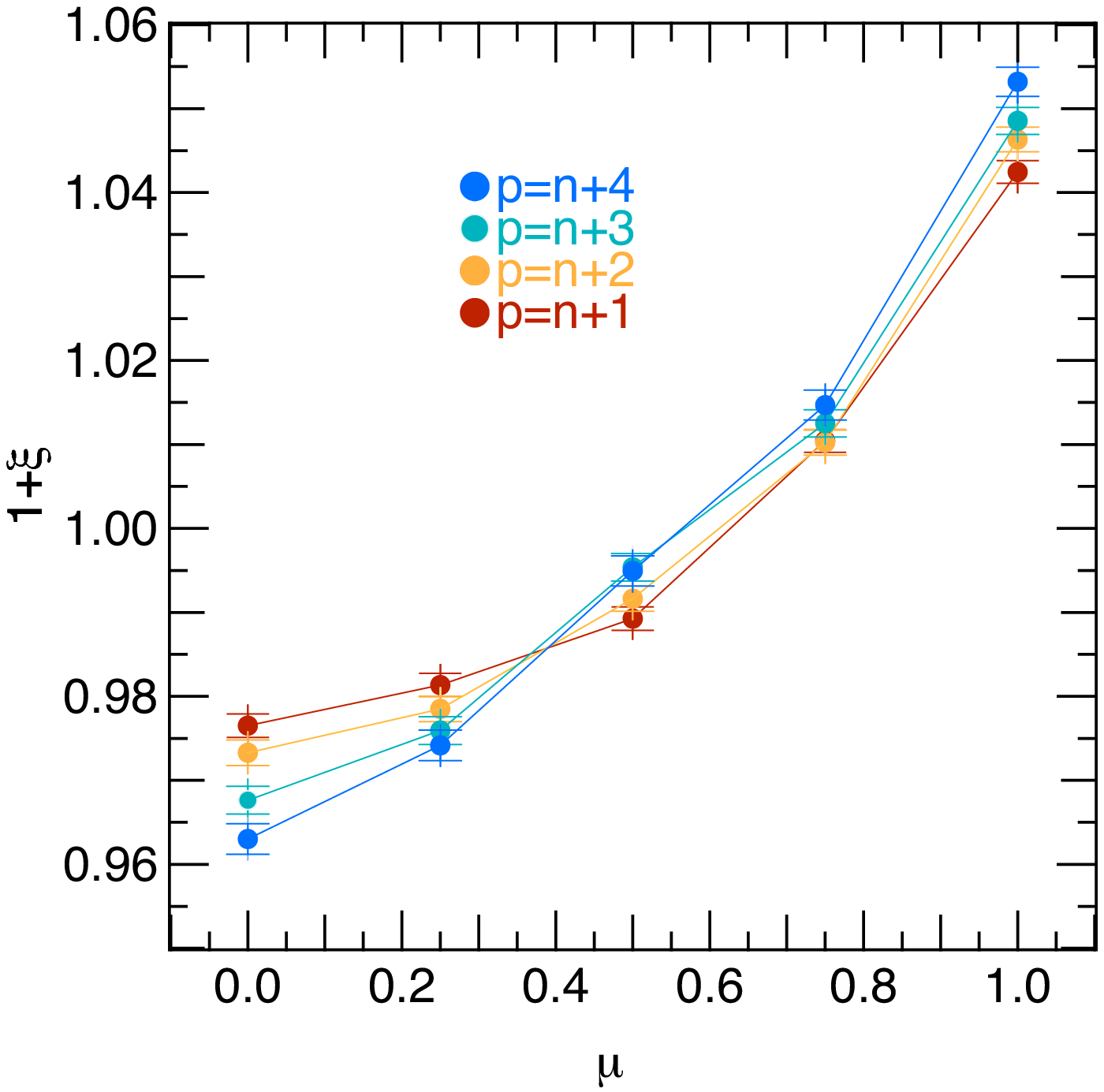}
      \includegraphics[width=3.7cm]{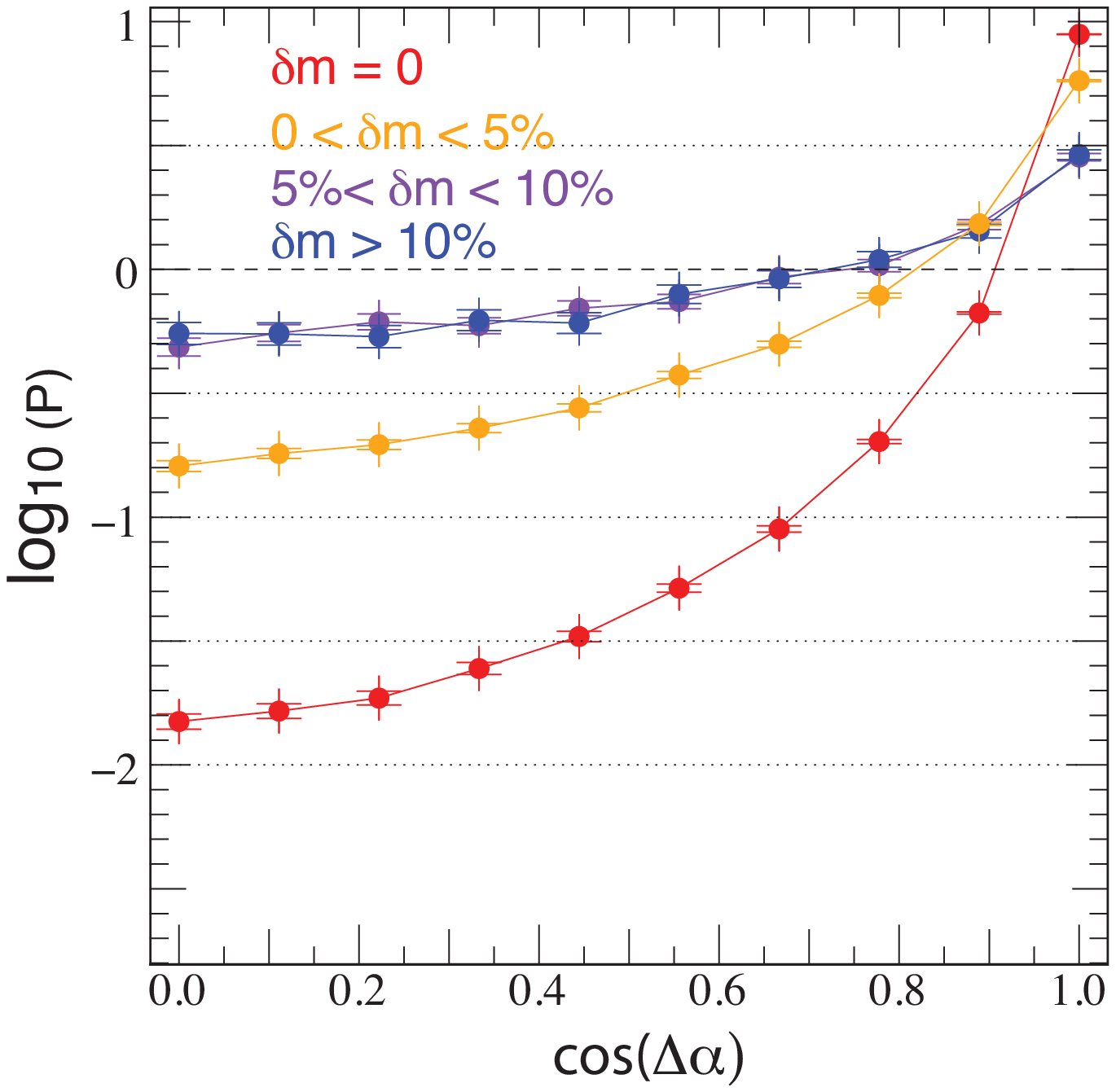} 
    \includegraphics[width=3.7cm]{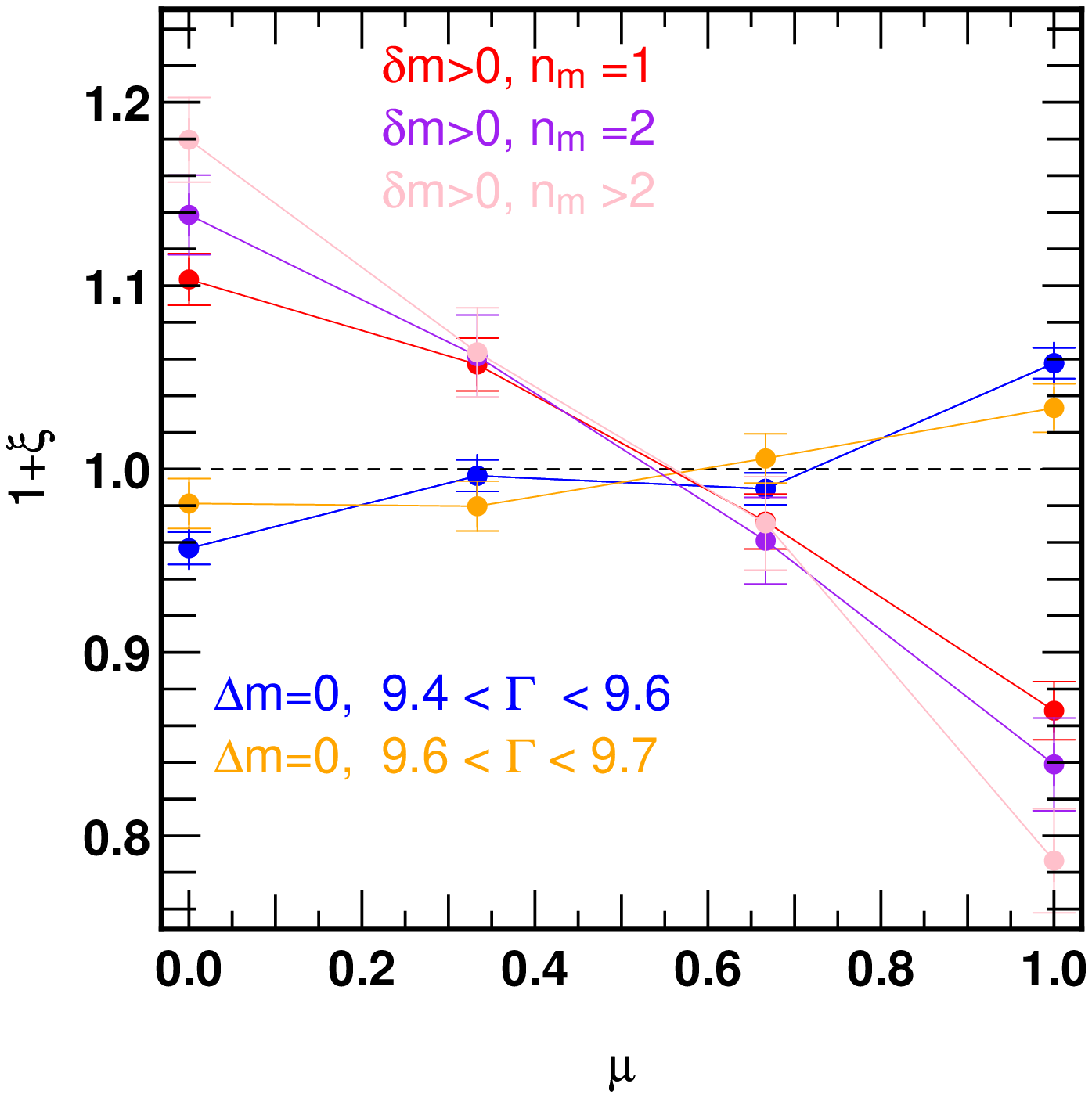}

 \caption{Panel (a) Excess probability, $\xi$, of $\mu$, the cosine of the angle between the spin of galaxies and their closest filament is shown for 4 consecutive outputs for galaxies which do not merge. Panel (b) the log(PDF) of  $\alpha$, the flip angle of the galactic angular momentum between two consecutive outputs is shown as a function of the merger ratio. Panel (c) same as (a) but as a function of the number of mergers $n_{m}$ the galaxy has undergone. $\Delta m =0$, absence of mergers across the history of the galaxy. $\Gamma= \log(M_{\rm}/M_{\odot})$.
Half sigma error bars are shown for readability. Dashed line is uniform PDF (excess probability $\xi=0$)}.
   \label{fig2}
\end{center}
\end{figure}

We argue that small galaxies with $M_{s}<M_{tr,s}$ grow their spin through smooth accretion (possibly from cold gas inflows) from the vorticity quadrant they are caught in. In the process, vorticity is transferred to the intrinsic angular momentum of the galaxy. This results in the (re)-alignment of the spin of those galaxies with their surrounding filament over time.

This can be seen on Fig.~\ref{fig2}(a) which displays the PDF of $\mu$ -the cosine of the spin-filament angle- for galaxies which do not merge over four consecutive outputs (with an average time step of 250 Myr). This shows that the excess probability $\xi$ increases for values close to unity over cosmic time, indicating a tendency for those galaxies to align to their closest filament.

Similarly, we find an average increase in galactic angular momentum amplitude over cosmic time for non-mergers. This confirms that smooth accretion tends to build the galactic spin over time. 

Finally, computing the inertia tensor of each galaxy from stellar kinematics and assuming an ellipsoidal shape, we recover the axis lengths and analyze the cumulative probabilities of the axis ratios over cosmic time. We find clear indication that smooth accretion flattens galaxies along this axis over cosmic time.

All findings sustain the  above detailed scenario. While it is easy to understand why the aligned component of the spin might not persist  when galaxies grow larger than their embedding vorticity quadrant, this does not explain why a clear perpendicular orientation is found for older massive galaxies.
 This leads us to investigate the effect of mergers.

{\underline{\it Spin ``anti''-alignments and mergers}}

Galaxies merge preferentially along  filaments. 
%
%
A pair of galaxies about to merge -- which are catching up relative to each other -- are more likely to display a relative orbital momentum perpendicular to the surrounding filament. As this momentum is converted into intrinsic angular momentum during the merger, the remnant is more likely to display a spin perpendicular to that filament. Moreover,  galaxies grow in mass from successive mergers. This was confirmed in the Horizon-AGN  simulation where we  found a clear correlation between merger fraction and stellar mass. Indeed, massive ellipticals are much more likely to be the result of mergers, which motivate their tendency to grow an orthogonal spin that will persist, and will not re-align from accretion providing that the galaxy is large enough.

To support this proposition, Fig.~\ref{fig2}(b) shows the log(PDF) of the spin flip angle between successive outputs for different merger ratios. If all the spins in the sample maintain a fixed orientation, we would expect a Dirac function centered on one, while if all the spins get randomly reinitialized at each time step, we expect the uniform PDF (the dashed line). On this figure, we see clearly that non-mergers  maintain their spin orientation (91 percent within a 25 degrees cone) while mergers trigger important swings, with a stronger effect for most massive mergers. This  strongly supports  the merger-driven swings scenario.

Fig.~\ref{fig2}(c) confirms this and displays the PDF of $\mu$ (as defined above) for different merger histories: we identify  an important excess probability $\xi$ of perpendicular orientation ($\mu$ close to 0) for mergers, with a increased amplitude for higher number of mergers across the history of the galaxy. Moreover the magnitude of this signal is at least comparable to that found for dark haloes in Codis et al. and three times stronger than the signals obtained for tracers in Dubois et al. Consistently, non-mergers tend to remain aligned to their closest filament.

\section{Conclusion}
\begin{itemize}
\item{ Smooth accretion from the surrounding vorticity quadrant builds up the spin of young galaxies parallel to the filament}
\item{ Mergers along the cosmic web 
 flip spins perpendicular to the filament through an orbital-to-intrinsic angular momentum transfer.} 
\item{ Galactic properties correlated to merger rate  trace the galactic flips and allow for detection.}
\item{ The scenario detailed here is the eulerian counterpart of the lagrangian theory (anisotropic TTT) presented by Christophe Pichon in the same proceedings, and shows how their predictions pervade down to small scales and low redshifts where non-linear baryonic physics become important.}
\end{itemize}

{\it This work is partially supported by the grant  ANR-13-BS05-0005.}

\end{document}